\documentclass[12pt,preprint]{aastex}

\begin{document}


\title{PRE-FLARE ACTIVITY AND MAGNETIC RECONNECTION DURING THE EVOLUTIONARY STAGES OF ENERGY RELEASE IN A SOLAR ERUPTIVE FLARE} 

\author{BHUWAN JOSHI\altaffilmark{1}, ASTRID M. VERONIG\altaffilmark{2}, JEONGWOO LEE\altaffilmark{3}, SU-CHAN BONG\altaffilmark{4}, SANJIV KUMAR TIWARI\altaffilmark{5}, AND KYUNG-SUK CHO\altaffilmark{4} } 

\altaffiltext{1}{Udaipur Solar Observatory, Physical Research Laboratory, Udaipur 313 001, India}
\altaffiltext{2}{IGAM/Institute of Physics, University of Graz, Universit$\ddot{a}$tsplatz 5, A-8010 Graz, Austria}
\altaffiltext{3}{Physics Department, New Jersey Institute of Technology, 161 Warren Street, Newark, NJ 07102, USA}
\altaffiltext{4}{Korea Astronomy and Space Science Institute, Daejeon 305-348, Korea}
\altaffiltext{5}{Max-Planck-Institut f\"{u}r Sonnensystemforschung, Max-Planck-Str. 2,
37191 Katlenburg-Lindau, Germany}

\begin{abstract}
In this paper, we present a multi-wavelength analysis of an eruptive white-light M3.2 flare which occurred in active region NOAA 10486 on November 1, 2003. Excellent set of high resolution observations made by RHESSI and TRACE provide clear evidence of significant pre-flare activities for $\sim$9 minutes in the form of an initiation phase observed at EUV/UV wavelengths followed by the X-ray precursor phase. During the initiation phase, we observed localized brightenings in the highly sheared core region close to the filament and interactions among short EUV loops overlying the filament which led to the opening of magnetic field lines. The X-ray precursor phase is manifested in RHESSI measurements below $\sim$30 keV and coincided with the beginning of flux emergence at the flaring location along with early signatures of the eruption. From the RHESSI observations, we conclude that both plasma heating and electron acceleration occurred during the precursor phase. The main flare is consistent with the standard flare model. However, after the impulsive phase, intense HXR looptop source was observed without significant footpoint emission. More intriguingly, for a brief period the looptop source exhibited strong HXR emission with energies up to 100 keV and significant non-thermal characteristics. The present study indicates a causal relation between the activities in the preflare and main flare. We also conclude that pre-flare activities, occurred in the form of subtle magnetic reorganization along with localized magnetic reconnection, played a crucial role in destabilizing the active region filament leading to solar eruptive flare and associated large-scale phenomena.

\end{abstract}

\keywords{Sun: corona --- Sun: flares --- Sun: X-rays}

\section{INTRODUCTION}
Solar eruptive phenomena correspond to various kinds of transient magnetic activities occurring in the solar atmosphere in the form of flares, eruptive prominences and coronal mass ejections (CMEs). With the availability of multi-wavelength data, especially from the space based platforms, it has become apparent that these are different manifestations of a single physical process implicating the disruption of coronal magnetic fields \citep[see e.g.,][]{lin03}. There is also a near-universal consensus that magnetic reconnection plays a key role in the process of disruption of magnetic fields as well as dissipation of stored magnetic energy in the corona \citep{lakhina2000,priest02}. 

Flares mostly occur in closed magnetic field configuration associated with active regions. Such closed magnetic structures may embrace one or more neutral lines in the photospheric magnetic flux. In a simplistic model, we can imagine the structure of a bipolar magnetic configuration in terms of an inner region, called the core field, and the outer region, called the envelope field \citep{moore2001}. The core fields are rooted close to the neutral line while the envelope fields are rooted away from it. Before an eruption, the core fields can usually be traced by a dark filament in the chromosphere when viewed on the solar disk. The core fields are usually strongly nonpotential in the pre-flare phase. 
In the initial stages of a large eruptive flare, the core fields containing the prominence erupt stretching out the envelope fields. With the evolution of the eruption process, the stretched field lines reclose via magnetic reconnection beneath the erupting filament. 
Multi-wavelength observations of solar eruptive flares have revealed several key features of the eruption process: rising arcade of intense (newly formed) soft X-ray loops, HXR and H$\alpha$ emission from the feet of the newly formed loops, and X-ray coronal source at their summit. The standard CSHKP model of solar flares has been successful to broadly incorporate these multi-wavelength flare components \citep[for a review see][]{hudson04,benz08,schrijver09}. 

Although the standard flare model successfully describes several observational features of a large eruptive flare, the basic question about the triggering of the eruption remains unclear and debatable. The two representative solar eruption models -- tether-cutting and breakout -- exploits the role of initial magnetic reconnection in two different ways in order to set up the conditions favorable for the core fields to erupt. 
The ``tether-cutting model" is fundamentally based on a single, highly sheared magnetic bipole, with the earliest reconnection occurring deep in the sheared core region \citep{moore2001}. On the other hand, in the ``breakout model" the
fundamental topology of the erupting system is multi-polar. Here the eruption is initiated by reconnection at a neutral point located in the corona, well above the core region \citep{antiochos99}. In this manner, the former is built on the concept of ``internal reconnection" while the latter is suggestive of an ``external reconnection" \citep{sterling2001}.

In order to understand the triggering mechanism of solar eruption, it is essential to examine the pre-eruption phase and probe those features which might have played vital role in the subsequent processes leading to fast energy release and eruption. Observations of solar flares in soft X-rays clearly indicate an enhancement in the flux before the flare, known as the X-ray precursor phase \citep{tappin1991}. 
There are evidences of active pre-flare structure in soft X--rays co-spatial with the main flare which develops several minutes or more before the onset of flare \citep{farnik96, farnik98,kim2008}.
However, we should not ignore the fact that significant pre-flare activities may  be present even before the X--ray precursor phase in other longer wavelength observations such as H$\alpha$ and EUV/UV. It has been suggested that the pre-flare brightening may occur as a result of slow reconnection and provide a trigger for the subsequent eruption \citep{moore92,chifor07}. Here it is worth mentioning that this pre-eruption reconnection may be very different from the post-eruption coronal reconnection, which is believed to lead a two-ribbon flare \citep{kim2001}.

In this paper, we present a comprehensive multi-wavelength analysis of a well observed M3.2 flare that occurred on November 1, 2003. The motivation of the present investigation is two fold: (1) to study the pre-flare activities and their relation with the eruption process, and (2) to understand the role of magnetic reconnection in the corona in the post-eruption phase. The study utilizes the excellent data sets from three space missions: Reuven Ramaty High Energy Solar Spectroscopic Imager (RHESSI), Transition Region and Coronal Explorer (TRACE) and Solar and Heliospheric Observatory (SOHO). These observations were supplemented by H$\alpha$ and vector magnetic measurements from ground based stations. Images of high temporal and spatial resolution at UV and EUV wavelengths coupled with 1-minute cadence SOHO/MDI magnetograms have enabled us to look for the minute changes that took place in the pre-flare phase. RHESSI X-ray imaging and spectroscopic analysis was performed to understand the thermal and non-thermal characteristics of the flare emission. In section 2, we present the data analysis and describe the multi-wavelength view of the event. In section 3, we integrate and discuss the observations presented in the previous section. The conclusions of the present study are summarized in section 4. 


\section{OBSERVATIONS AND DATA ANALYSIS}

\subsection{Event overview}

\label{sec_lc}
The Active region NOAA 10486 (along with 10484 and 10488) produced several powerful eruptions during the period of October-November 2003. According to the solar region summary reports compiled by the Space Weather Prediction Center\footnote{http://www.swpc.noaa.gov/} (SWPC) AR 10486 appeared between 23 October and 5 November, 2003. The observations presented here correspond to the flare activity that occurred in AR 10486 on November 1, 2003 at the location S12 W60 showing GOES SXR intensity of M3.2 and H$\alpha$ class of 1N.

According to GOES reports the flare took place between 22:26 and 22:49 UT with a peak at 22:38 UT. In Figure \ref{event_lc}, we provide the GOES lightcurves in 0.5--4 and 1--8~\AA~wavelength bands. Figure \ref{ha_mdi_rhessi} provides multi-wavelength aspects of the active region. In Figure \ref{ha_mdi_rhessi}a we show a representative H$\alpha$ filtergram on November 1, 2003 obtained from the Udaipur Solar Observatory (USO). We find a filament at the north-west part of the active region (marked by an arrow) which was partially erupted during the flare. A comparison of H$\alpha$ filtergram with TRACE white light (WL) image (Figure \ref{ha_mdi_rhessi}b) shows three sunspots in the vicinity of the location where the filament erupted, two at the eastern side and one at the western side of it. It is evident that the two sunspots at the eastern side are more prominent. In Figures \ref{ha_mdi_rhessi}c \& d, we plot a TRACE WL image and a SOHO/MDI magnetogram respectively during the flare main phase overlaid by RHESSI X-ray images. 


The event was completely observed by RHESSI spacecraft \citep{lin02}. 
In order to understand the X-ray emission during the eruption process, we have analyzed the X-ray lightcurves and images (see section \ref{sec_imaging}) in 
four energy bands namely 6-12, 12-25, 25-50, and 50-100 keV. The RHESSI light curves, shown in Figure \ref{event_lc}, are constructed by taking average count rates over front detectors 1, 3--6, 8, and 9 in each energy band. We note several important aspects of variation in X-ray fluxes that indicate important stages of the flare evolution: (1) X-ray count rates at 6--12 and 12--25 keV energy bands show a bump at $\sim$22:26 UT. At this time the X-ray flux at high energy bands ($\gtrsim$ 25 keV) is still at the background level. This bump indicates the precursor phase of the flare and is the most prominent in the 12--25 keV energy band. RHESSI 12--25 keV light curve clearly describes the X--ray precursor phase between 22:24 and 22:28 UT. (2) GOES as well as RHESSI time profiles show a steady rise in the flux from $\sim$22:28 UT which indicates the beginning of the flare impulsive phase. We observe a peak at $\sim$22:30 UT simultaneously in the three energy channels (12--25, 25--50, and 50--100 keV). This peak appears even sharper as we consider light curves of higher energy bands. This peak cannot be recognized in the GOES time profiles and RHESSI lightcurve in 6--12 keV energy band. Hereafter we describe this peak as the first HXR burst. (3) The emission at high energy bands at energies $\gtrsim$ 25 keV further enhances after 22:31 UT and a second HXR burst is observed at $\sim$22:33 UT. (4) In low energy bands, there is a gradual rise in the count rates and the light curves peak several minutes after the second HXR burst. A third HXR burst occurred during the decline phase at $\sim$22:39 UT. Table~\ref{tab1} presents a summary of different phases of the flare evolution.


\subsection{Multi-wavelength imaging}
\label{sec_imaging}

\subsubsection{X-ray and E(UV) observations}
The RHESSI images have been reconstructed with the CLEAN algorithm with the natural weighing scheme using front detector segments 3 to 8 (excluding 7) in different
energy bands, namely, 6--12, 12--25, 25--50, and 50--100 keV \citep{hurford02}. We compare RHESSI measurements with TRACE images in 195~{\AA} and 1600~{\AA}
wavelengths. The TRACE 195~{\AA} filter is mainly sensitive to plasmas at a temperature around 1.5 MK (Fe XII) but during flares it may also contain significant contributions of plasmas at temperatures around 15--20 MK \citep[due to an Fe XXIV line;][]{handy99}. The TRACE 1600~{\AA} channel is sensitive to plasma in the temperature range between (4--10)$\times$10$^{3}$ K and represent a combination of UV continuum, C I, and Fe II lines \citep{handy99}. Also the brightest and most rapidly varying features in TRACE 1600~{\AA} channel are likely to emit in the C IV lines \citep{handy98}.

It is known that the pointing of TRACE is not very accurate. Therefore, in order to compare RHESSI images and TRACE images, we need to correct the pointing information of TRACE images. This is achieved by considering the fact that the pointing information of RHESSI and SOHO is quite accurate. Therefore, we corrected TRACE pointing by comparing a TRACE WL and a SOHO WL image observed at 22:23:29 and 22:23:33 UT, respectively. For the cross correlation, we used the Solar SoftWare (SSW) routine, trace\_mdi\_align, developed by T. Metcalf \citep[see also][]{metcalf03}.

We first describe the observations of TRACE in the 195~{\AA} EUV channel along with co-temporal RHESSI X-ray images in Figure \ref{trace195_rhessi}. TRACE images reveal a dark, elongated, inverted U-shaped structure that already existed in the flaring region which corresponds to the filament (this filament is marked in H$\alpha$ image shown in Figure \ref{ha_mdi_rhessi}a). 
The pre-flare 195~{\AA} images reveal that the filament was thinner at the middle (cf. Figure \ref{trace195_rhessi}a). Near the middle of the elongated filamentary structure, we observe very first signatures of the flare in the form of brightenings at both sides of the filament material at $\sim$22:19 UT. It is noteworthy that this initial EUV flare brightening was observed $\sim$4 minutes before the start of X-ray precursor phase. We call these two bright sources as northwest (NW) and southeast (SE) EUV kernels (indicated by red arrows Figure \ref{trace195_rhessi}b). The TRACE EUV movie during 22:19--22:24 UT reveals rapid changes in the configuration of short loops and interactions among them in a region (indicated in Figure \ref{trace195_rhessi}a and b by white arrows) which is confined over the west part of the filament. A careful examination of EUV images between $\sim$22:21--22:23 UT clearly reveals that two closely situated short loops show rapid structural changes, forming a cusp-like feature. At $\sim$22:26 UT, this region brightens up and an arc-like bright structure is formed that expanded rapidly in the direction of the cusp (indicated by an arrow in Figure \ref{trace195_rhessi}c). We therefore interpret the rapid evolution of filed lines into the cusp shaped structure to represent stretching or opening of magnetic field system.
It is important to note that this very moment, $\sim$22:26 UT, coincides with the peak of the X-ray precursor phase (recognized during 22:24--22:28 UT; cf. subsection \ref{sec_lc} and Figure \ref{event_lc}). We observe first clear X-ray sources at this time at energies below 25 keV (cf. Figure \ref{trace195_rhessi}c). 
The arc-like EUV structure and the two EUV kernels continued to become brighter and thicker. We note that the arc-like structure appears the most prominent and intense in the image at 22:29:58 UT which corresponds to the first HXR burst (cf. Figure \ref{event_lc}). The next image at 22:31:23 UT, just after the first HXR burst, represents very crucial stage of the flare evolution. To further emphasize the EUV observation at and just after the first HXR burst, we provide a clearer and closer view of the activity site in Figure \ref{trace195} which distinctly reveals restructuring of the flaring region. The comparison of the two images show several interesting new developments: the arc-like structure becomes fainter and moves toward southwest (marked by white arrows), a new intense source develops close to the NW kernels (marked by blue arrow), bright patches of emission are seen between the new source and the arc-like structure (marked by yellow arrow), and a bright loop system is observed (marked by red arrow). At the eastern part of the loop system, a group of bright points appear which are likely to represent the emission from the footpoints of the loop system. The next image available to us is after a gap of $\sim$6 min which shows bright and closed system of loops having simplified structure (cf. Figure \ref{trace195_rhessi}i). The brightness of the closed loop system decays slowly and the loops become more structured in the later stages.  

In Figure \ref{trace1600_rhessi}, we present the UV observations of the flare taken in the TRACE 1600~{\AA} passband along with co-temporal RHESSI X-ray images. The careful examination of UV images suggests that the initial flare brightening occurred at $\sim$22:19 UT in the form of two bright kernels, very similar to that of EUV measurements. However, within 2 minutes, the two kernels rapidly evolve into ribbon-like structures (indicated by arrows in Figure \ref{trace1600_rhessi}a). We call these ribbons as northwest (NW) and southeast (SE) ribbons. The most noticeable feature observed in UV images at the early stage is the brightening at a relatively remote location (marked by an arrow in Figure \ref{trace1600_rhessi}b), close to the southeast sunspot (cf. Figure \ref{ha_mdi_rhessi}), which continues during the flare main phase. This brightening cannot be seen in EUV images. We find that this near-sunspot brightening is connected to the SE flare ribbon by a less-bright, thin, elongated structure along which transient bright points appear. It is noteworthy that in RHESSI 12--25 keV image we observe X-ray emission from this remote bright region for a short period at around 22:28 UT (cf. Figure \ref{trace1600_rhessi}e) which mark the beginning of the flare impulsive phase (cf. Figure \ref{event_lc} and subsection \ref{sec_lc}). In UV images we observe sporadic brightening during $\sim$22:20--22:25 UT over a small region (at western portion of images) where EUV images show rapid changes in the configuration of short loops overlying the filament. However, after $\sim$22:25 UT we see the formation of the arc-like structure, consistent with the flare evolution in EUV wavelength, around the location of sporadic brightening (marked by an arrow in Figure {\ref{trace1600_rhessi}c). From UV and EUV observations described till now, we collect several important pieces of information that clearly reveal an initiation phase of the flare between $\sim$22:19 and 22:24 UT, before the flare signatures in X-ray observations.      

Near the first HXR burst ($\sim$22:30 UT) we observe two strong 50--100 keV X-ray sources that lie over the two UV flare ribbons. At this time we also find that a loop-like structure evolves that connects the two UV flare ribbon. The intensity and thickness of the loop increases rapidly. We mark the loop system by an arrow in Figure \ref{trace1600_rhessi}h. During the second HXR burst ($\sim$22:33 UT) we again find two high energy X-ray sources at 50--100 keV. 
The association of UV flare ribbons with the strong HXR sources suggests that the two HXR sources mark the conjugate footpoints (FP) of a loop system.
The HXR emissions from the footpoints of flaring loops are traditionally   viewed in terms of the thick-target bremsstrahlung process \citep{brown71} in which the X-ray production at the FPs of the loop system takes place when high-energy electrons, accelerated in the reconnection region, come along
the guiding magnetic field lines and penetrate the denser transition
region and chromospheric layers \citep[cf.][]{kontar2010}.
We observe a low energy X-ray source (below 25 keV), below the erupting arc-like structure, throughout the flare. We, therefore, interpret that the low energy X-ray source indicates the region of X-ray emission from the top of the hot loops.

In Figure \ref{rhessi_images}, we present RHESSI images in 6--12 keV, 12--25 keV, 25--50 keV, and 50--100 keV energy bands to show the temporal and spatial evolution of X-ray sources. We find that the origin and evolution of the X-ray sources during the precursor phase is very interesting and requires careful examination. Since the X-ray time profiles indicate significant emission below 30 keV (cf. Figures \ref{event_lc} and \ref{spec_param}d), we investigate the morphology of series of images in 6--12 keV and 12--25 keV during this phase. Although the X-ray flux starts to rise at $\sim$22:24 UT, the clear structure of X-ray sources was observed from 22:26 UT onward, i.e., from the peak of X-ray precursor phase. The evolution of X-ray sources between 22:26:00 UT and 22:28:30 UT reveals several interesting features. At the beginning the emission at 6--12 keV originate in the form of X-ray flare ribbons (cf. Figure \ref{rhessi_images}a) which are formed at both sides of the filament (Figure \ref{trace195_rhessi}c). The next image at 22:26:30 UT shows similar morphology although now ribbons contract and rather resemble footpoint sources. Here it is noteworthy that the X-ray source at high energy, i.e. 12--25 keV, lies in between 6--12 keV X-ray ribbons/footpoints (cf. Figure \ref{rhessi_images}b). The next images seem to show a looptop but again the footpoint emission dominate. This scenario is further supported by comparing X-ray source locations with TRACE UV images (cf. Figure \ref{trace1600_rhessi}e). The TRACE EUV images during this interval indicates a continuous evolving bright feature (termed earlier as arc-like structure) above the filament. Therefore we find that the multi-wavelength analysis of the initiation and precursor phases provides clear evidence of subtle activity and magnetic reorganization before the flare onset.


Around the first and second HXR bursts, the X-ray emission in 25--50 keV and 50--100 keV energy bands  originates from two distinct FP sources. It is to be noted that the separation between the two FPs around the second HXR burst is less than that of the first HXR bursts (Figure \ref{rhessi_images}). On the other hand, emission at energies below 25 keV is confined to a single looptop (LT) source that slowly moves towards southwest side of the activity site. It is noteworthy that from $\sim$22:34 UT onwards, X-ray emission at 25-50 keV energy band also originates from the location of the LT source and we can observe single X-ray source simultaneously in 6-12 keV, 12--25 keV, 25--50 keV energy bands. Between $\sim$22:34 and $\sim$22:38 UT, the HXR flux in 50--100 keV energy band becomes very weak and we cannot see source structure. However, at $\sim$22:39 UT HXR flux enhances (cf. Figure \ref{event_lc}) and a strong HXR source appears which is co-spatial with the LT source observed at low energies (see Figure \ref{rhessi_images}p). In Figure \ref{rhessi_LT_FP}, we plot the altitude evolution of RHESSI LT source in different energy bands as well as the separation of the two 50--100 keV FP sources. For 6--12, 12--25, and 25--50 keV measurements, the LT altitude is defined as the distance along the main axis of motion between the centroid of the LT source and a reference point on the Sun (the center of the line between the two FPs seen in RHESSI 50--100 keV image at 22:29 UT). The LT source is defined as the region with emission above 85\% of the peak flux of each image. In Figure {\ref{rhessi_LT_FP}, we also plot the height of 50--100 keV LT source which is seen only for a short period during the third HXR burst at $\sim$22:39 UT. In this case, we determine the LT source by selecting a region with emission above 75\% of the peak flux.

\subsubsection{X-ray and WL observations}
This flare also produced signatures in white light (WL) observations. We examined series of TRACE WL images which are available typically every minute. The WL channel of TRACE has a very broad response, from 1700~{\AA} to 1 $\mu$m making it sensitive to emission in transition region, chromosphere, and photosphere \citep{handy99}. We find flare related brightening in WL images from $\sim$22:29 UT that lasted for $\sim$7 minutes. Figure \ref{trace_wl} shows a few representative WL images. These images clearly reveal that brightenings occurred at three locations: WL ribbon very close to western sunspot which is co-spatial with the NW~HXR source (one of the FPs), another WL ribbon close to eastern sunspot which is co-spatial with the SE~HXR source (second HXR FP), and a bright patch of emission visible only for $\sim$2 minutes between the two HXR bursts (indicated by an arrow in Figure \ref{trace_wl}c). 
The fact that HXR FP sources correlate well with the WL brightenings suggests this event to be a WL flare of type I \citep{fang95}.
We also notice that initially northern edge of the NW WL ribbon touched the sunspot while in the later stages ribbon separated from the sunspot.    


\subsection{RHESSI X-ray spectroscopy}
\label{sec_spec}

We have studied the evolution of RHESSI X-ray spectra during the flare over consecutive 20 s intervals from the precursor to the decline phase (i.e., between 22:25:00 UT and 22:40:00 UT). For this analysis we first generated a RHESSI spectrogram with an energy binning of $\frac{1}{3}$ keV from 6--15 keV and 1 keV from 15--80 keV. We only used the front segments of the detectors, and excluded detectors 2 and 7 (which have lower energy resolution and high threshold energies, respectively). The spectra were deconvolved with the full detector response matrix \citep[i.e., off-diagonal elements were included;][]{smith02}. 

In Figure \ref{rhessi_spec} we show spatially integrated, background subtracted  RHESSI spectra derived during six time intervals of the flare together with the applied spectral fits. This six time intervals are marked in Figure \ref{spec_param}d where we have plotted 6--30 keV and 30--80 keV RHESSI light curves. The plot reveals that during the precursor phase emission originates only at low X--ray energies (i.e, 6--30 keV). We find rapid increase in X-ray flux in 30--80 keV energy band only with the onset of impulsive phase at $\sim$22:28 UT. Therefore, we have restricted the spectral fitting during the precursor phase (between 22:25:00 and 22:28:20 UT) in the energy range 6--30 keV. The spectra derived after the impulsive phase (between 22:28:20 and 22:40:00 UT) were fitted in the range 6--80 keV.

Spectral fits were obtained using a forward-fitting method implemented in the OSPEX code. The OSPEX allows the user to choose a model photon spectrum, which is multiplied with the instrument response matrix and then fitted to the observed cound spectrum. The best-fit parameters are obtained as output. 
We used the bremsstrahlung and line spectrum of an isothermal plasma, and a power-law function with a turn-over at low X-ray energies. The negative power-law index below the low energy turn-over was fixed at 1.5. In this manner, there are five free parameters in the model: temperature (T) and emission measure (EM) for the thermal component; power-law index ($\gamma$), normalization of the power-law, and low energy turn-over for the non-thermal component. From these fits, we derive the temperature and emission measure of the hot flaring plasma as well as the power law index for the non-thermal component. Figure {\ref{spec_param} shows the time evolution of these parameters obtained from fits to the RHESSI spectra integrated over consecutive 20 s intervals.

The spectra during the precursor phase suggest that hot thermal emission with temperature T $>$ 24 MK already existed at the very start (cf. Figure \ref{spec_param}a). The comparison of RHESSI images with UV and EUV observations indicates that this intense plasma heating corresponds to localized brightenings at three regions in the form of two X-ray ribbons/footpoints along with a looptop source. However, the temperature decreases afterward for a short period (between 22:26:50 and 22:27:50 UT). The temperature again increases with the onset of the impulsive phase (after 22:27:50 UT) and peaks ($\sim$30 MK) at 22:28:50 UT. The temperature does not increase any further in the later stages. 
The emission measure EM shows a gradual increase during the precursor phase (between 22:25:10 and 22:27:50 UT) followed by a decrease for a short duration. The EM further shows a gradual rise with the start of the impulsive phase until the end of the HXR emission. Around the peak of the precursor phase, we observe significant HXR emission. During this time interval, the spectra at energies $\varepsilon \gtrsim 10$ keV can be fitted by a power law with photon spectral index $\gamma$ in the range of $\sim$6--7.


The spectra reveal quite different characteristics during the impulsive and decay phase. During the period $\sim$22:29--22:34 UT, when the high energy HXR emission originates from FPs, the spectra follow hard power laws with photon spectral index $\gamma$ in the range of $\sim$4--5. The hardest X-ray emission is produced around the second HXR burst at 22:33:30 UT with $\gamma = 3.6$. It is interesting to note that there is again a significant increase in the HXR emission at a later stage of the flare at $\sim$22:39 UT. Around this interval also the spectral index is harder with $\gamma =3.9$ at 22:38:50 UT (cf. Figures \ref{rhessi_spec}f and \ref{spec_param}). It is rather unusual that at this interval the HXR emission solely originates from the LT location (cf. Figure \ref{rhessi_images}p), due to the low density in the corona which makes the bremsstrahlung process inefficient.

\subsection{Magnetic structure of the flaring region}
\label{sec_mag_structure}
In order to understand the magnetic configuration of the activity site and its role in driving the eruptive phenomenon we analyze SOHO/MDI data \citep{scherrer95}. In Figure \ref{mdi_wl_mag}a we show a SOHO white light pre-flare image in which we have marked different sunspots near the flaring region as S1, S2, and S3. As described in subsection \ref{sec_lc}, S1 and S2 are located toward the eastern side of the filament while S3 is at the western side of the filament. We have thoroughly examined the 1 minute cadence SOHO/MDI magnetogram movie before and during the flare between 21:00 UT and 23:00 UT. In Figure \ref{mdi_wl_mag} (panels b-d), we show a few representative magnetograms.  Regions of different magnetic polarities are indicated by arrows in Figure \ref{mdi_wl_mag}b. The comparison of panels (a) and (b) suggests that S1 and S2 are sunspots of negative polarity while S3 is a positive polarity sunspot. 
The sequence of magnetogram images of the activity site reveals the emergence of magnetic flux in a region close to sunspot S3. This region is marked in Figure \ref{mdi_wl_mag}d as EFR. From the Figure it is apparent that the EFR is of positive polarity. However, we cannot be certain of the intrinsic magnetic polarity of the EFR because of its location at W60. In Figure \ref{mdi_flux}, we plot emerging magnetic flux through EFR by selecting a rectangular box of size 28$''$ $\times$ 20$''$ (cf. Figure \ref{mdi_wl_mag}d). The important observation is that the magnetic flux of the EFR starts to rise in one polarity at $\sim$22:25 UT which coincides with the onset of X-ray precursor phase.


In Figure \ref{mdi_wl_mag}c, we show the location of X-ray emission during the flare impulsive phase over the co-temporal magnetogram. We find the NW and SE hard X-ray FPs are associated with the positive polarity region (S3) and negative polarity region (F-) respectively. 

In Figure \ref{msfc_vector}, we show a vector magnetogram of the active region NOAA 10486 on November 1, 2003 at 14:08 UT. The magnetogram data has been taken from the vector magnetograph facility of Marshall Space Flight Center \citep{hagyard82}. The 180$^{o}$ azimuthal ambiguity has been resolved by using minimum energy method \citep{metcalf94,leka09}. The transverse vector fields are indicated by green arrows in the figure. 
The region associated with flaring activity is shown inside the red box in Figure \ref{msfc_vector}. The direction of transverse vectors between the negative polarity regions (associated with sunspots S1 and S2) and positive polarity region (associated with sunspot S3) indicate that magnetic field lines are highly sheared. To quantify the non-potentiality of the field lines in the activity site, we compute the spatially averaged signed shear angle \citep[SASSA;][]{tiwari09} which represents the average deviation of the observed transverse vectors from that of the potential transverse vectors. Over the region of interest (described by red box in Figure \ref{msfc_vector}) we obtain the value of SASSA as $\sim$15$^{o}$. Such a high value of SASSA indicates that the region was highly stressed, and capable of driving major eruptions \citep{tiwari10}.

\section{RESULTS AND DISCUSSION}

We divide the whole flare activity into four distinct evolutionary stages and discuss important characteristic features of each phase.



\subsection{Initiation phase ($\sim$22:19--22:24 UT)}

The initiation phase is readily visible at UV and EUV wavelengths. UV images indicate the initial brightenings at two locations, one on each side of the filament. This brightening looks much like flare kernels. It is noteworthy that main flare occurred at this location only. EUV images reveal rapid changes in the configuration of short loops embedding the filament. 
We observed interactions among short EUV loops which resulted in the opening of field lines. Vector magnetogram and EUV images suggest that the magnetic field is highly sheared at this site. We further notice that the discrete, localized brightenings can be identified in both UV and EUV images. 

It is evident that the initiation phase represents the initial energy release at distinct locations in a region of of highly sheared magnetic fields close to the filament, i.e., the ``core" of the erupting region. It is likely that during this phase a small volume of plasma is heated up at different locations insufficient to produce detectable level of X-ray emission. 


\subsection{Precursor phase ($\sim$22:24--22:28 UT)}
The precursor phase shows significant X-ray emissions below 30 keV, while the counts rates at higher energies ($\geq$30 keV) are still at the background level. We find an emerging flux region (EFR) within the core region, and very first signatures of the eruption in the form of a bright arc-like feature in UV/EUV images. Since an EFR may destabilize the sheared magnetic structures leading to solar eruption \citep{choudhary98}, it is likely that the onset of eruption is intimately connected to the emergence of magnetic flux. Further, the initial eruption took place at the location where EUV loops interacted during the initiation phase. 

The plasma temperature was very high at the beginning of the precursor phase, but the emission measure was still low and increased gradually (cf. Figure \ref{spec_param}). This indicates that the low energy X-rays emission at this stage is originated from discrete volumes of hot plasma. This fact is further confirmed with EUV/UV images which still display localized brightenings which are cospatial with the X-ray sources and corresponds to emission from X-ray ribbons/footpoints and looptop. We find that around the peak of the precursor phase HXR emission follows power law which provides evidence for electron acceleration. 

To synthesize the initiation and precursor phases, the initial energy release took place in the form of localized brightenings in the highly sheared core region and is associated with the early signatures of eruption. It has been suggested that the X-ray precursor phase, with distinct, localized brightenings can be understood in terms of localized magnetic reconnection which acts as a common trigger for both flare emission and filament eruption \citep{chifor07}. 

    
\subsection{Impulsive phase ($\sim$22:28--22:34 UT)} 

The impulsive phase is represented by the onset of high energy ($\geq$ 30 keV) HXR emission at 22:28 UT indicating the impulsive release of large amount of energy. The plasma temperature also rises impulsively and attains a maximum value of $\sim$30 MK at 22:28:50 UT. The temperature slowly decreases in the later stages throughout the flare while emission measure gradually increases. It implies that now the flare involves a larger volume with the filling of hot plasma in the loop system \citep{uddin03}. 

At the time of maximum plasma temperature, the EUV images show a major re-organization in the structure of the flaring region as indicated in Figure \ref{trace195}. 
The multi-wavelength signatures observed at HXR, UV, and WL measurements at this stage show consistency with the standard flare model \citep [see, e.g.,][]{joshi07,joshi2009}. Spatial correlation between HXR FP sources and WL emitting regions suggests that WL emission is closely connected with the flare energy deposition by non-thermal particles in the chromosphere \citep{hudson72,metcalf03}.


It is important to note that the converging motion of HXR FPs at the beginning of the impulsive phase is marked by rapidly evolving EUV loops into a simplified structure, indicating a possible connection between the two processes. We interpret this as an evidence for the relaxation of highly sheared magnetic loops \citep{ji07,joshi2009}.

The X-ray spectra exhibit a significant non-thermal component throughout the impulsive phase with the hardest X-ray emission at the time of the second HXR burst. In general, we note a distinct anti-correlation between the evolution of HXR flux and photon spectral index (cf. Figure \ref{spec_param}c \& d). Such a behavior indicates that each non-thermal emission peak represents a distinct acceleration event of the electrons in the flare \citep{grigis04}.


\subsection{Decay phase ($\sim$22:34--22:49 UT)} 
During this interval GOES soft X-rays attain the maximum phase. We found a HXR source at 25--50 keV at the top of the EUV flare loop system throughout the decline phase which shows continuous upward motion, consistent with the standard flare model. However, it is noteworthy that  the HXR looptop source is observed without footpoint component.
High energy HXR LT source is believed to be closely associated with the site of electron acceleration in the corona \citep{krucker07,krucker08,krucker10}. 
Moreover, we observed a HXR burst in this late phase, at $\sim$22:39 UT, during which spectrum shows significant non-thermal emission with $\gamma =$ 3.9. At this time a single HXR source was detected for a brief period which is cospatial with the 25--50 keV LT source and shows movement in the same direction. Further it is located away from the HXR FP sources detected earlier during the impulsive phase (cf. Figure \ref{rhessi_images}). Therefore, we interpret that HXR emission at 50--100 keV at this time originates from the looptop. 
Coronal HXR emission has been reported in some of the recent RHESSI observations \citep{lin2003,veronig2004,veronig2005,krucker08a,krucker08b}. However, the physical mechanism for such a strong non-thermal source in the tenuous corona is still not clearly understood. Here it is very interesting to see that the strong HXR emission from the LT source between 22:34 and 22:46 UT is temporally associated with the steep rise in the magnetic flux emergence (cf. Figure \ref{mdi_flux}). Further we find that the LT source seems to be spatially located within the EFR region (Figure \ref{mdi_wl_mag}). 
Therefore it is likely that the new magnetic flux was continuously fed to the magnetic reconnection site in the corona causing the prolonged non-thermal looptop emission.

\section{CONCLUSIONS}

The availability of excellent high-cadence multi-wavelength data has enabled us to make a detailed investigation of the physical processes that led to the M3.2 flare on November 1, 2003 and the associated eruption. The main emphasis of this study lies in understanding the pre-flare activity which manifested for $\sim$9 minutes before the onset of flare impulsive phase. The early preflare activities are characterized in the form of an initiation phase, recognized in EUV and UV wavelengths, which is followed by more energetic X-ray precursor phase. 
The main activity during the initiation phase is the localized brightenings at three locations close to a filament in a highly sheared magnetic field region. The main flare showed emissions exactly at the same locations. Another important observation of the initiation phase is the rapid changes in the configuration of short EUV loops followed by opening of field lines. The very first signature of eruption was seen during X-ray precursor phase at the location where EUV loops interacted.  

The onset of the X-ray precursor phase coincided with the flux emergence. The X-ray precursor phase is characterized by high plasma temperatures, with maximum temperature $\sim$28 MK, and corresponding EUV/UV images showed enhanced brightening along with plasma eruption. More importantly, we find HXR non-thermal emission which suggests that electron acceleration occurred during the precursor phase. We therefore conclude that pre-flare brightenings correspond to events of localized magnetic reconnection in the core region, i.e., close to neutral line where filament lies. It is likely that the interactions among short EUV loops, overlying the filament, followed by the flux emergence played a crucial role in driving the eruption and successive large-scale magnetic reconnection that resulted the main flare.

The impulsive phase of the flare is mostly consistent with the standard flare scenario. However, a HXR looptop source is observed during the impulsive as well as decay phase. It is noteworthy that HXR LT source became stronger in the decay phase and showed non-thermal emission. Further, the HXR LT sources at the decay phase is rather unusual in that there is no significant footpoint emission.

The present study indicates a causal relation between pre-flare activity and main flare. It also follows that the signatures of magnetic reconnection during the initiation and precursor phase occur in the form of localized instances of energy release. In this manner, one can  differentiate pre-eruption reconnection from the post-eruption coronal reconnection which is generally understood in the framework of standard flare model. Our understanding of the pre-eruption reconnection is still limited because of observational constraints. However, we should keep in mind that some times the earliest pre-flare activities can be anticipated with EUV/UV measurements, well before the X-ray precursors. The new data sets from Solar Dynamic Observatory (SDO), with superior resolution, would be very useful for such investigations.

\acknowledgments
We sincerely acknowledge the anonymous referee for critical comments that provide a new direction to discuss the observational results and significantly improved the quality of the manuscript. We acknowledge RHESSI, TRACE, SOHO, and GOES for their open data policy. RHESSI and TRACE are NASA's small explorer missions. SOHO is a joint project of international cooperation between the ESA and NASA. We acknowledge the NASA/MSFC data archive for providing magnetogram data. AV gratefully acknowledges support by the European Community Framework Programme 7, ``High Energy Solar Physics Data in Europe (HESPE)", grant agreement no.: 263086.  J. L. was supported by NSF grant AST-0908344. This work was partially supported by the ``Development of Korean Space Weather Center" of KASI and the KASI basic research funds. We thank M. Karlick{\'y} for useful discussions.


\begin{thebibliography}{}

\bibitem[\protect\citeauthoryear{{Antiochos}, {DeVore}, \&
  {Klimchuk}}{{Antiochos} et~al.}{1999}]{antiochos99}
{Antiochos}, S.~K., {DeVore}, C.~R.,  \& {Klimchuk}, J.~A. 1999, \apj, 510, 485

\bibitem[\protect\citeauthoryear{{Benz}}{{Benz}}{2008}]{benz08}
{Benz}, A.~O. 2008, Living Reviews in Solar Physics, 5, 1

\bibitem[\protect\citeauthoryear{{Brown}}{{Brown}}{1971}]{brown71}
{Brown}, J.~C. 1971, \solphys, 18, 489

\bibitem[\protect\citeauthoryear{{Chifor} et~al.}{{Chifor}
  et~al.}{2007}]{chifor07}
{Chifor}, C., {Tripathi}, D., {Mason}, H.~E.,  \& {Dennis}, B.~R. 2007, \aap,
  472, 967

\bibitem[\protect\citeauthoryear{{Choudhary}, {Ambastha}, \& {Ai}}{{Choudhary}
  et~al.}{1998}]{choudhary98}
{Choudhary}, D.~P., {Ambastha}, A.,  \& {Ai}, G. 1998, \solphys, 179, 133

\bibitem[\protect\citeauthoryear{{Fang} \& {Ding}}{{Fang} \&
  {Ding}}{1995}]{fang95}
{Fang}, C.,  \& {Ding}, M.~D. 1995, \aaps, 110, 99

\bibitem[\protect\citeauthoryear{{F{\'a}rn{\'{\i}}k}, {Hudson}, \&
  {Watanabe}}{{F{\'a}rn{\'{\i}}k} et~al.}{1996}]{farnik96}
{F{\'a}rn{\'{\i}}k}, F., {Hudson}, H.,  \& {Watanabe}, T. 1996, \solphys, 165,
  169

\bibitem[\protect\citeauthoryear{{F{\'a}rn{\'{\i}}k} \&
  {Savy}}{{F{\'a}rn{\'{\i}}k} \& {Savy}}{1998}]{farnik98}
{F{\'a}rn{\'{\i}}k}, F.,  \& {Savy}, S.~K. 1998, \solphys, 183, 339

\bibitem[\protect\citeauthoryear{{Grigis} \& {Benz}}{{Grigis} \&
  {Benz}}{2004}]{grigis04}
{Grigis}, P.~C.,  \& {Benz}, A.~O. 2004, \aap, 426, 1093

\bibitem[\protect\citeauthoryear{{Hagyard} et~al.}{{Hagyard}
  et~al.}{1982}]{hagyard82}
{Hagyard}, M.~J., {Cumings}, N.~P., {West}, E.~A.,  \& {Smith}, J.~E. 1982,
  \solphys, 80, 33

\bibitem[\protect\citeauthoryear{{Handy} et~al.}{{Handy}
  et~al.}{1999}]{handy99}
{Handy}, B.~N., et~al. 1999, \solphys, 187, 229

\bibitem[\protect\citeauthoryear{{Handy} et~al.}{{Handy}
  et~al.}{1998}]{handy98}
{Handy}, B.~N., {Bruner}, M.~E., {Tarbell}, T.~D., {Title}, A.~M., {Wolfson},
  C.~J., {Laforge}, M.~J.,  \& {Oliver}, J.~J. 1998, \solphys, 183, 29

\bibitem[\protect\citeauthoryear{{Hudson} et~al.}{{Hudson}
  et~al.}{2004}]{hudson04}
{Hudson}, H., {Fletcher}, L., {Khan}, J.~I.,  \& {Kosugi}, T. 2004, in
  Astrophysics and Space Science Library, Vol. 314, Solar and Space Weather
  Radiophysics, ed. D.~E. {Gary} \& C.~{Keller}, 153

\bibitem[\protect\citeauthoryear{{Hudson}}{{Hudson}}{1972}]{hudson72}
{Hudson}, H.~S. 1972, \solphys, 24, 414

\bibitem[\protect\citeauthoryear{{Hurford} et~al.}{{Hurford}
  et~al.}{2002}]{hurford02}
{Hurford}, G.~J., et~al. 2002, \solphys, 210, 61

\bibitem[\protect\citeauthoryear{{Ji}, {Huang}, \& {Wang}}{{Ji}
  et~al.}{2007}]{ji07}
{Ji}, H., {Huang}, G.,  \& {Wang}, H. 2007, \apj, 660, 893

\bibitem[\protect\citeauthoryear{{Joshi} et~al.}{{Joshi}
  et~al.}{2007}]{joshi07}
{Joshi}, B., {Manoharan}, P.~K., {Veronig}, A.~M., {Pant}, P.,  \& {Pandey}, K.
  2007, \solphys, 242, 143

\bibitem[\protect\citeauthoryear{{Joshi} et~al.}{{Joshi}
  et~al.}{2009}]{joshi2009}
{Joshi}, B., et~al. 2009, \apj, 706, 1438

\bibitem[\protect\citeauthoryear{{Kim} et~al.}{{Kim} et~al.}{2001}]{kim2001}
{Kim}, J., {Yun}, H.~S., {Lee}, S., {Chae}, J., {Goode}, P.~R.,  \& {Wang}, H.
  2001, \apjl, 547, L85

\bibitem[\protect\citeauthoryear{{Kim} et~al.}{{Kim} et~al.}{2008}]{kim2008}
{Kim}, S., {Moon}, Y., {Kim}, Y., {Park}, Y., {Kim}, K., {Choe}, G.~S.,  \&
  {Kim}, K. 2008, \apj, 683, 510

\bibitem[\protect\citeauthoryear{{Kontar} et~al.}{{Kontar}
  et~al.}{2010}]{kontar2010}
{Kontar}, E.~P., {Hannah}, I.~G., {Jeffrey}, N.~L.~S.,  \& {Battaglia}, M.
  2010, \apj, 717, 250

\bibitem[\protect\citeauthoryear{{Krucker} et~al.}{{Krucker}
  et~al.}{2008a}]{krucker08}
{Krucker}, S., et~al. 2008a, \aapr, 16, 155

\bibitem[\protect\citeauthoryear{{Krucker}, {Hannah}, \& {Lin}}{{Krucker}
  et~al.}{2007}]{krucker07}
{Krucker}, S., {Hannah}, I.~G.,  \& {Lin}, R.~P. 2007, \apjl, 671, L193

\bibitem[\protect\citeauthoryear{{Krucker} et~al.}{{Krucker}
  et~al.}{2010}]{krucker10}
{Krucker}, S., {Hudson}, H.~S., {Glesener}, L., {White}, S.~M., {Masuda}, S.,
  {Wuelser}, J.,  \& {Lin}, R.~P. 2010, \apj, 714, 1108

\bibitem[\protect\citeauthoryear{{Krucker} et~al.}{{Krucker}
  et~al.}{2008b}]{krucker08a}
{Krucker}, S., {Hurford}, G.~J., {MacKinnon}, A.~L., {Shih}, A.~Y.,  \& {Lin},
  R.~P. 2008b, \apjl, 678, L63

\bibitem[\protect\citeauthoryear{{Krucker} \& {Lin}}{{Krucker} \&
  {Lin}}{2008}]{krucker08b}
{Krucker}, S.,  \& {Lin}, R.~P. 2008, \apj, 673, 1181

\bibitem[\protect\citeauthoryear{{Lakhina}}{{Lakhina}}{2000}]{lakhina2000}
{Lakhina}, G.~S. 2000, Bulletin of the Astronomical Society of India, 28, 593

\bibitem[\protect\citeauthoryear{{Leka}, {Barnes}, \& {Crouch}}{{Leka}
  et~al.}{2009}]{leka09}
{Leka}, K.~D., {Barnes}, G.,  \& {Crouch}, A. 2009, in Astronomical Society of
  the Pacific Conference Series, Vol. 415, Astronomical Society of the Pacific
  Conference Series, ed. {B.~Lites, M.~Cheung, T.~Magara, J.~Mariska, \&
  K.~Reeves}, 365

\bibitem[\protect\citeauthoryear{{Lin}, {Soon}, \& {Baliunas}}{{Lin}
  et~al.}{2003}]{lin03}
{Lin}, J., {Soon}, W.,  \& {Baliunas}, S.~L. 2003, \nar, 47, 53

\bibitem[\protect\citeauthoryear{{Lin} et~al.}{{Lin} et~al.}{2002}]{lin02}
{Lin}, R.~P., et~al. 2002, \solphys, 210, 3

\bibitem[\protect\citeauthoryear{{Lin} et~al.}{{Lin} et~al.}{2003}]{lin2003}
{Lin}, R.~P., et~al. 2003, \apjl, 595, L69

\bibitem[\protect\citeauthoryear{{Metcalf}}{{Metcalf}}{1994}]{metcalf94}
{Metcalf}, T.~R. 1994, \solphys, 155, 235

\bibitem[\protect\citeauthoryear{{Metcalf} et~al.}{{Metcalf}
  et~al.}{2003}]{metcalf03}
{Metcalf}, T.~R., {Alexander}, D., {Hudson}, H.~S.,  \& {Longcope}, D.~W. 2003,
  \apj, 595, 483

\bibitem[\protect\citeauthoryear{{Moore} \& {Roumeliotis}}{{Moore} \&
  {Roumeliotis}}{1992}]{moore92}
{Moore}, R.~L.,  \& {Roumeliotis}, G. 1992, in Lecture Notes in Physics, Berlin
  Springer Verlag, Vol. 399, IAU Colloq. 133: Eruptive Solar Flares, ed.
  {Z.~Svestka, B.~V.~Jackson, \& M.~E.~Machado}, 69

\bibitem[\protect\citeauthoryear{{Moore} et~al.}{{Moore}
  et~al.}{2001}]{moore2001}
{Moore}, R.~L., {Sterling}, A.~C., {Hudson}, H.~S.,  \& {Lemen}, J.~R. 2001,
  \apj, 552, 833

\bibitem[\protect\citeauthoryear{{Priest} \& {Forbes}}{{Priest} \&
  {Forbes}}{2002}]{priest02}
{Priest}, E.~R.,  \& {Forbes}, T.~G. 2002, \aapr, 10, 313

\bibitem[\protect\citeauthoryear{{Scherrer} et~al.}{{Scherrer}
  et~al.}{1995}]{scherrer95}
{Scherrer}, P.~H., et~al. 1995, \solphys, 162, 129

\bibitem[\protect\citeauthoryear{{Schrijver}}{{Schrijver}}{2009}]{schrijver09}
{Schrijver}, C.~J. 2009, Advances in Space Research, 43, 739

\bibitem[\protect\citeauthoryear{{Smith} et~al.}{{Smith}
  et~al.}{2002}]{smith02}
{Smith}, D.~M., et~al. 2002, \solphys, 210, 33

\bibitem[\protect\citeauthoryear{{Sterling} et~al.}{{Sterling}
  et~al.}{2001}]{sterling2001}
{Sterling}, A.~C., {Moore}, R.~L., {Qiu}, J.,  \& {Wang}, H. 2001, \apj, 561,
  1116

\bibitem[\protect\citeauthoryear{{Tappin}}{{Tappin}}{1991}]{tappin1991}
{Tappin}, S.~J. 1991, \aaps, 87, 277

\bibitem[\protect\citeauthoryear{{Tiwari}, {Venkatakrishnan}, \&
  {Gosain}}{{Tiwari} et~al.}{2010}]{tiwari10}
{Tiwari}, S.~K., {Venkatakrishnan}, P.,  \& {Gosain}, S. 2010, \apj, 721, 622

\bibitem[\protect\citeauthoryear{{Tiwari}, {Venkatakrishnan}, \&
  {Sankarasubramanian}}{{Tiwari} et~al.}{2009}]{tiwari09}
{Tiwari}, S.~K., {Venkatakrishnan}, P.,  \& {Sankarasubramanian}, K. 2009,
  \apjl, 702, L133

\bibitem[\protect\citeauthoryear{{Uddin} et~al.}{{Uddin}
  et~al.}{2003}]{uddin03}
{Uddin}, W., {Joshi}, B., {Chandra}, R.,  \& {Joshi}, A. 2003, Bulletin of the
  Astronomical Society of India, 31, 303

\bibitem[\protect\citeauthoryear{{Veronig} \& {Brown}}{{Veronig} \&
  {Brown}}{2004}]{veronig2004}
{Veronig}, A.~M.,  \& {Brown}, J.~C. 2004, \apjl, 603, L117

\bibitem[\protect\citeauthoryear{{Veronig}, {Brown}, \& {Bone}}{{Veronig}
  et~al.}{2005}]{veronig2005}
{Veronig}, A.~M., {Brown}, J.~C.,  \& {Bone}, L. 2005, Advances in Space
  Research, 35, 1683

\end{thebibliography}

\clearpage

\begin{table}
\begin{center}
\caption{Summary of different phases of the flare evolution}
\begin{tabular}{cccc}
\tableline\tableline
Phases  & Start time--End time (UT) & Observing wavelength\\
\tableline
 Initiation phase & 22:19--22:24  & EUV and UV  \\
 Precursor phase & 22:24--22:28  & X-ray ($\lesssim 30$ keV), EUV, and UV \\
 Impulsive Phase & 22:28--22:34  & X-ray (upto $\sim$100 keV), EUV, and UV  \\
 Decay phase & 22:34--22:49  &  X-ray (upto $\sim$100 keV), EUV, and UV \\
\tableline
\label{tab1}
\end{tabular}
\end{center}
\end{table}

\begin{figure}
\epsscale{.80}
\plotone{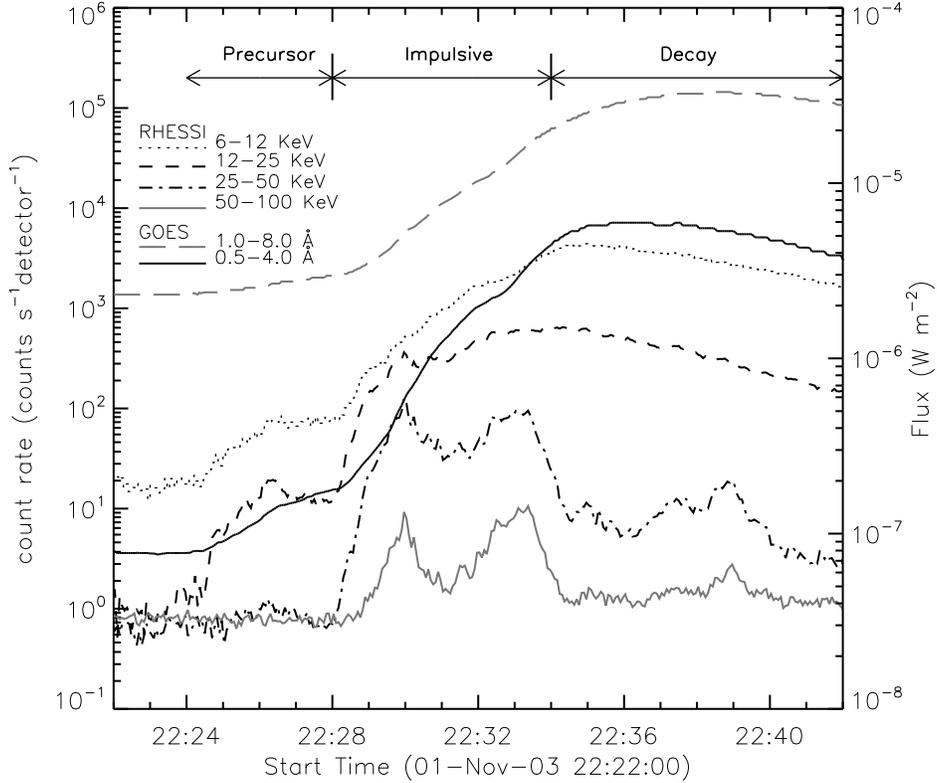}
\caption{RHESSI and GOES lightcurves of the flare with a time cadence of 4 s and 3 s respectively. In order to present different RHESSI light curves with clarity, the RHESSI count rates are scaled by factors of 1, 1/4, 1/5, and 1/10 for the energy bands 6--12, 12--25, 25--50, and 50--100 keV, respectively.}
\label{event_lc}
\end{figure}

\begin{figure}
\epsscale{.80}
\plotone{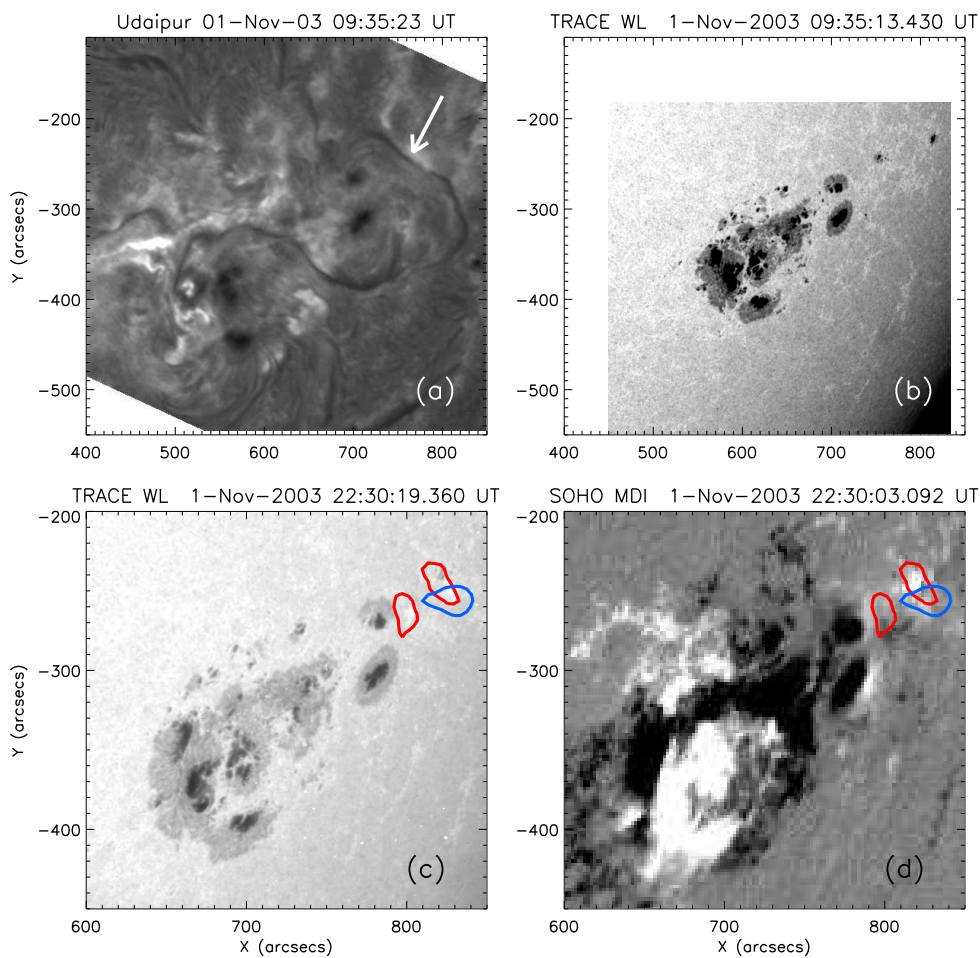}
\caption{Multi-wavelength view of the active region during pre-flare and flare timings. (a) H$\alpha$ filtergram taken from Udaipur Solar Observatory at about $\sim$13 hours before the event. The filament that erupted during the flare is marked by an arrow. (b) White light observation of the active region by TRACE taken at the time of H$\alpha$ filtergram shown in panel (a). We find that the filament lie in the north-west part of the active region. (c) \& (d) TRACE white light image and SOHO/MDI magnetogram during the impulsive phase of the flare. Red and blue contours represent co-temporal X-ray sources at 50--100 keV and 6--12 keV energy bands respectively and denote the region where the X-ray intensity is 60\% of its peak value.}
\label{ha_mdi_rhessi}
\end{figure}

\begin{figure}
\plotone{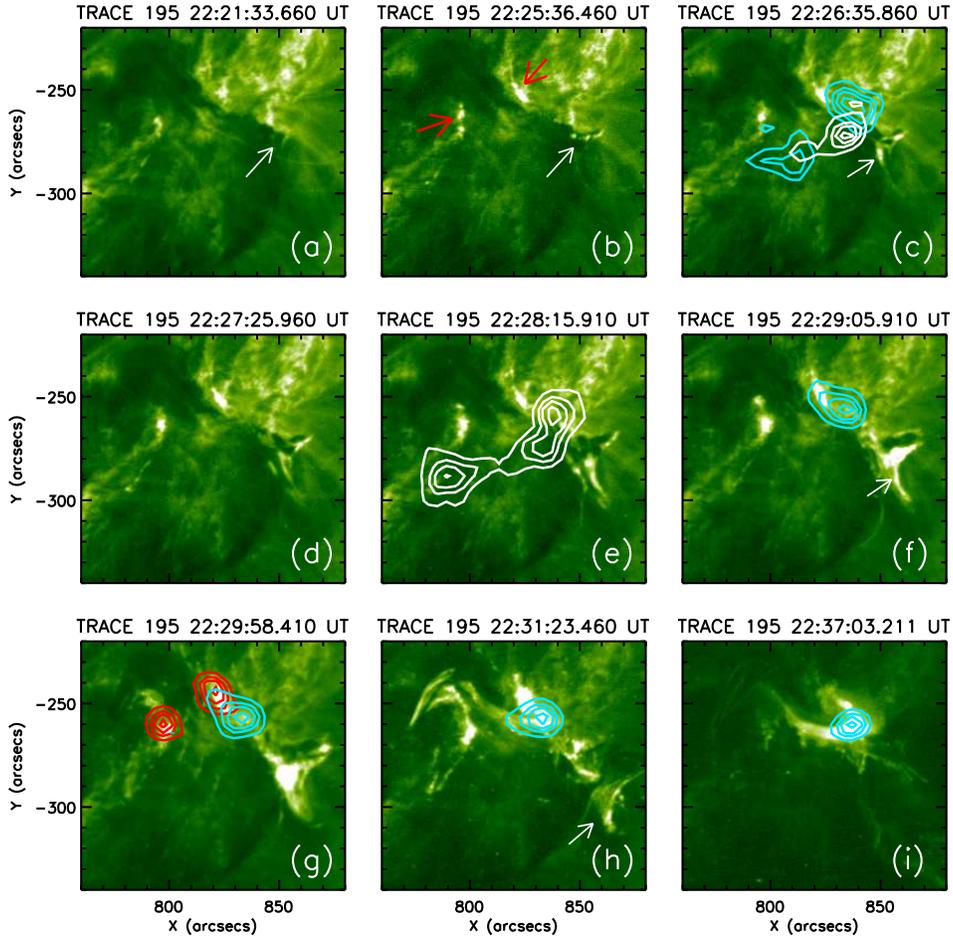}
\caption{Sequence of TRACE 195~\AA~images from pre-flare to post-flare stages. Panels c, and e--i show co-temporal RHESSI X-ray images in 6--12 keV (blue), 12--25 keV (white) and 50--100 keV (red) energy bands overlaid on TRACE images. RHESSI images are reconstructed with the CLEAN algorithm using grids 3--8 and natural weighing scheme. The integration time for RHESSI images is 30 s. The contour levels for RHESSI images are 60\%, 75\%, 85\%, and 95\% of the peak flux in each image.}
\label{trace195_rhessi}
\end{figure}

\begin{figure}
\epsscale{.50}
\plotone{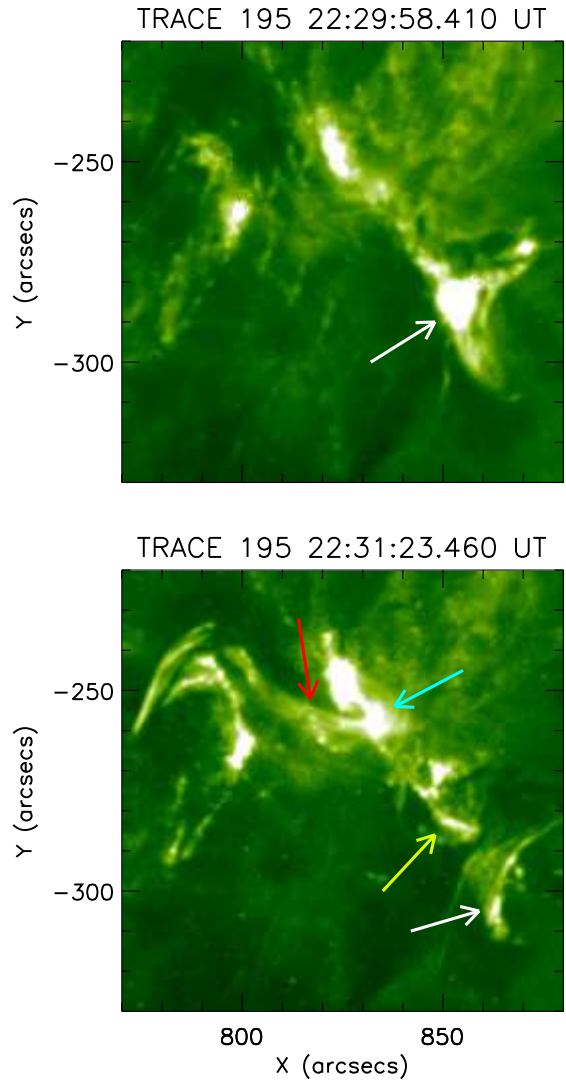}
\caption{TRACE 195~\AA~images just before and after the first HXR burst.}
\label{trace195}
\end{figure}

\begin{figure}
\epsscale{.80}
\plotone{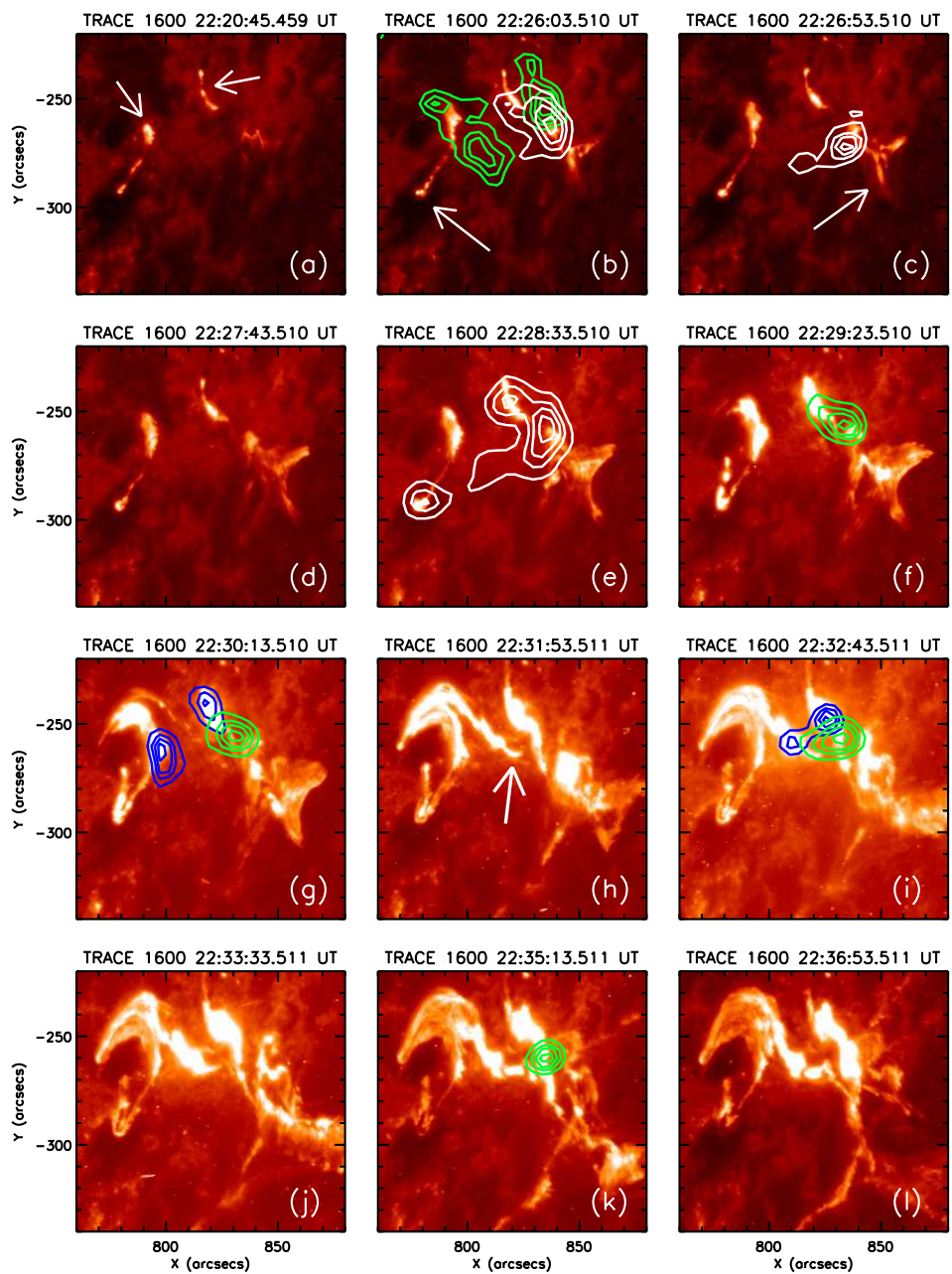}
\caption{Sequence of TRACE 1600~\AA~images from pre-flare to post-flare stages. Panels b-c, e--g, i, and k show co-temporal RHESSI X-ray images in 6--12 keV (green), 12--25 keV (white) and 50--100 keV (blue) energy bands overlaid on TRACE images. The contour levels for RHESSI images are 60\%, 75\%, 85\%, and 95\% of the peak flux in each image. RHESSI image parameters are the same as in Figure \ref{trace195_rhessi}.}
\label{trace1600_rhessi}
\end{figure}

\begin{figure}
\epsscale{.80}
\plotone{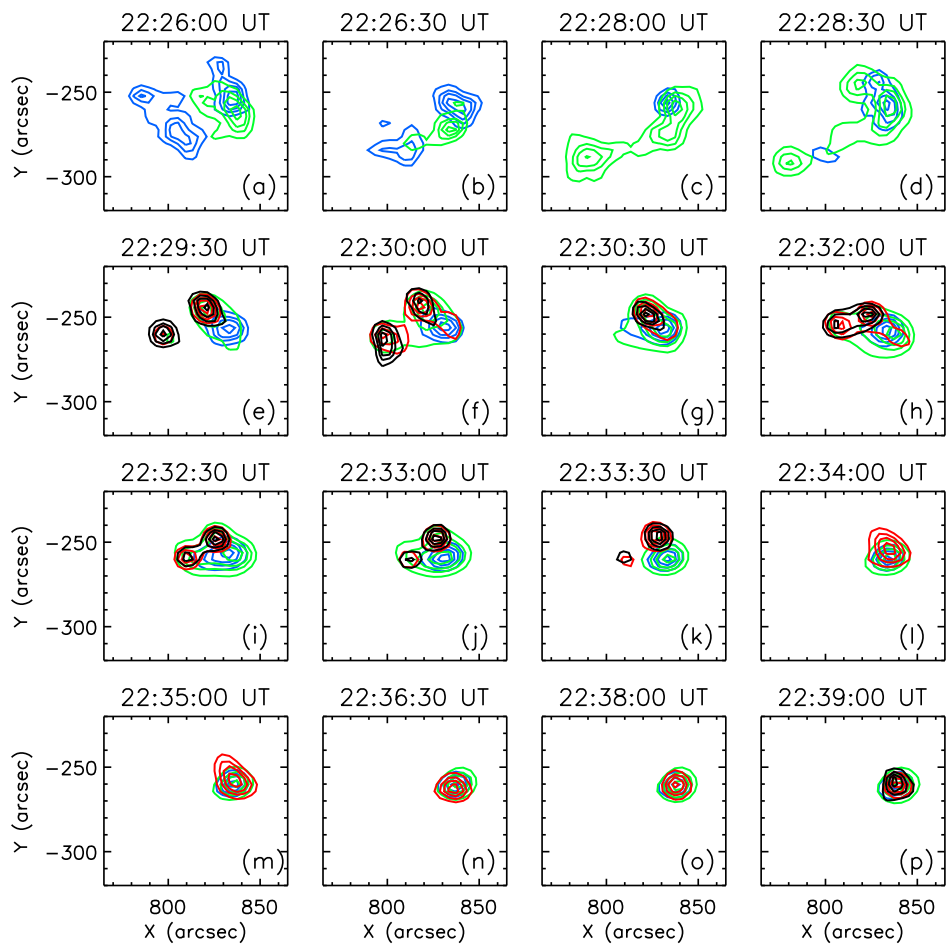}
\caption{Temporal evolution of X-ray sources in the 6--12 keV (blue), 12--25 keV (green), 25--50 keV (red) and 50--100 keV (black) energy bands. The contour levels for RHESSI images are 60\%, 75\%, 85\%, and 95\% of the peak flux in each image. RHESSI image parameters are the same as in Figure \ref{trace195_rhessi}.}
\label{rhessi_images}
\end{figure}

\begin{figure}
\epsscale{.80}
\plotone{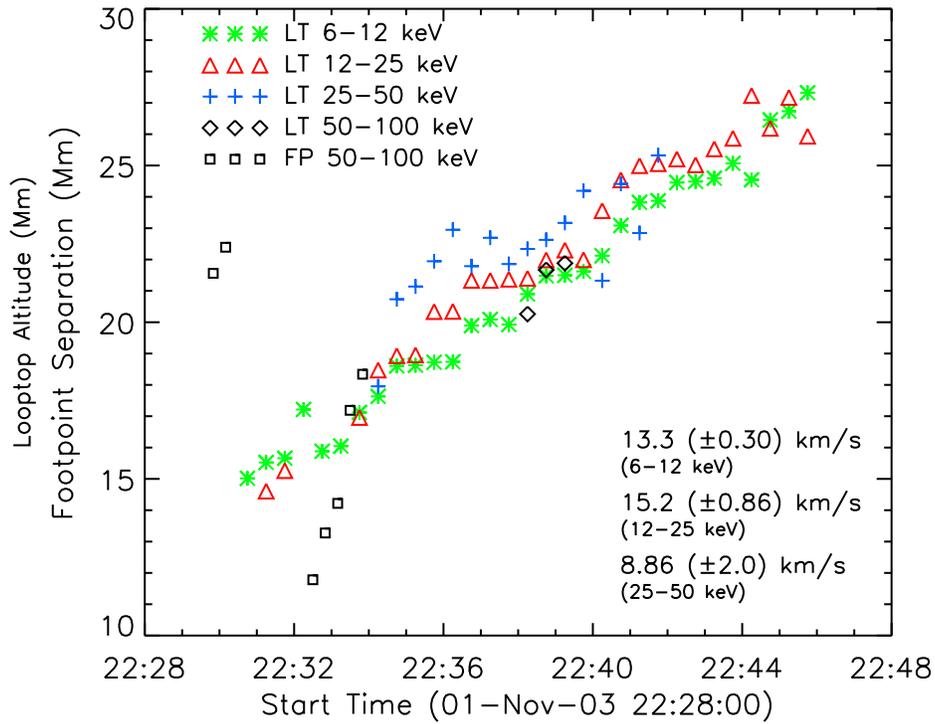}
\caption{Evolution of the altitude of the RHESSI looptop source observed in the 6--12, 12--25, 25--50, and 50--100 keV energy bands. Note the 50--100 keV LT source appeared only for a short period (three data points at $\sim$22:39 UT). Additionally, the   footpoint separation derived from RHESSI 50--100 keV images is plotted.}
\label{rhessi_LT_FP}
\end{figure}

\begin{figure}
\epsscale{.80}
\plotone{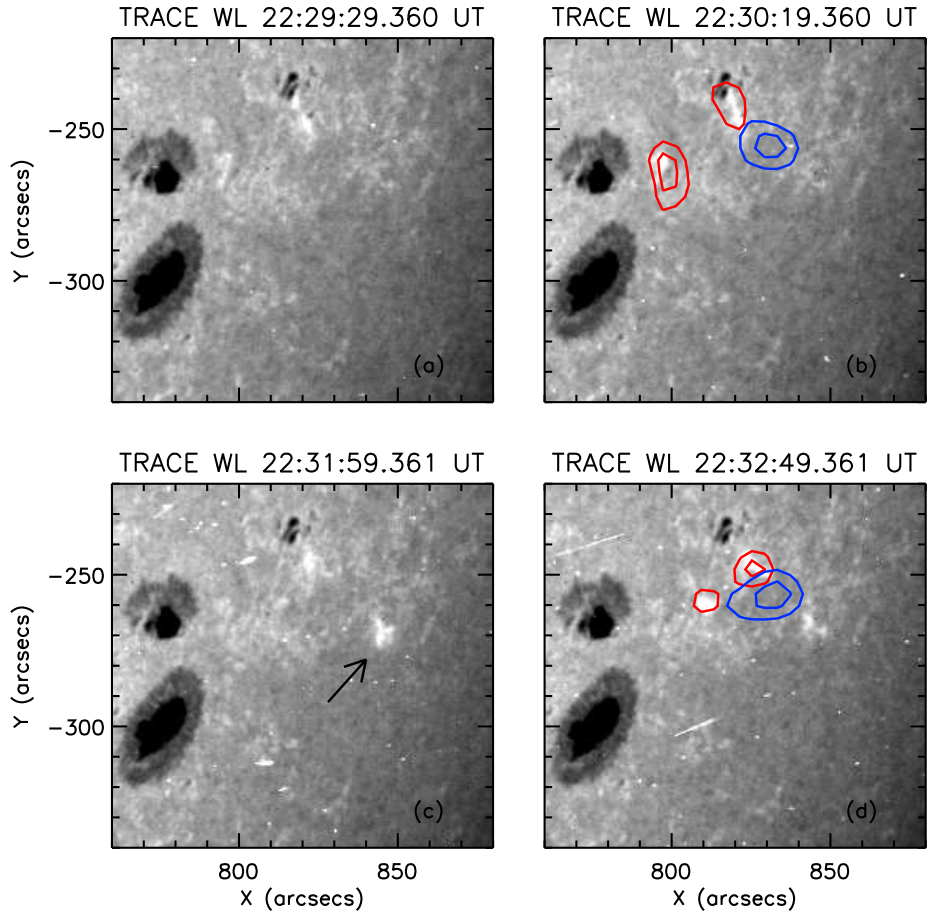}
\caption{Sequence of  TRACE WL images during the flare. Panels b and d also show contours of co-temporal X-ray sources at 6--12 keV (blue) 50--10 keV (red) energy bands. The contour levels for RHESSI images are 70\% and 90\% of the peak flux in each image.}
\label{trace_wl}
\end{figure}


\begin{figure}
\epsscale{.80}
\plotone{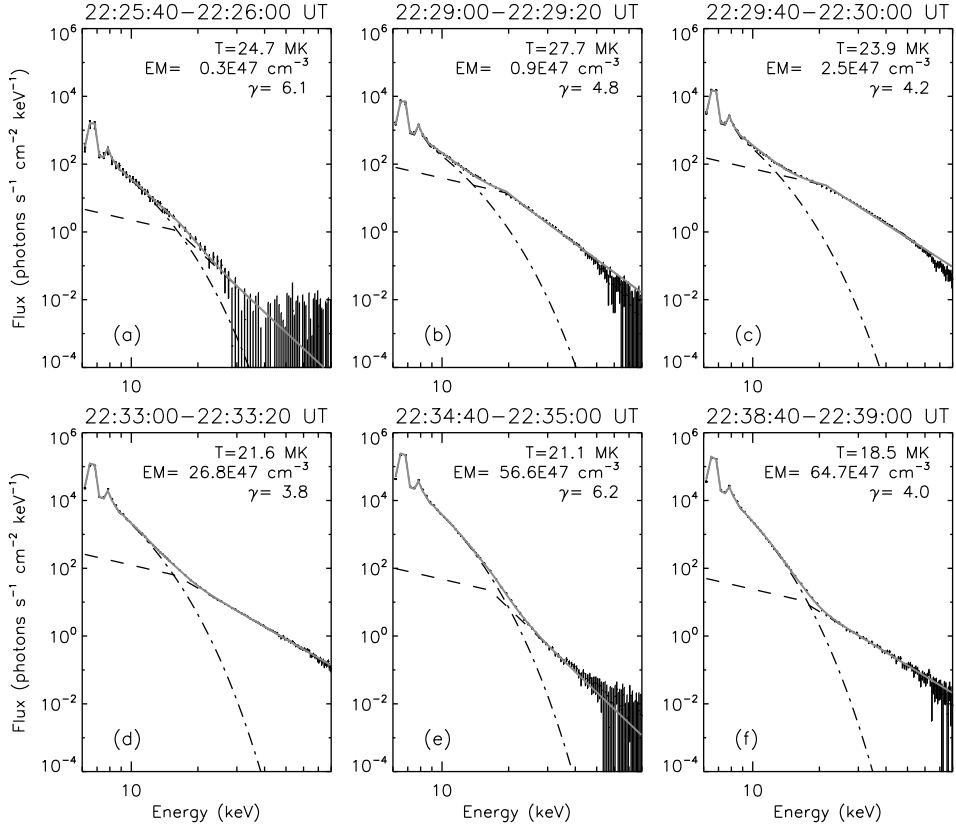}
\caption{RHESSI X-ray spectra derived during six selected time intervals during the flare (indicated in Figure \ref{spec_param} by vertical lines) together with applied fits. The spectra were fitted with an isothermal  
model (dashed-dotted line) and a functional power law with a turn-over at low energies (dashed line). The gray (solid) line indicates the sum of the two components. The spectrum during the precursor phase (panel a) was fitted in the energy range 6--30 keV, while the spectra derived during the main phase (panels b-f) were fitted in the range 6--80 keV.}
\label{rhessi_spec}
\end{figure}

\begin{figure}
\epsscale{.80}
\plotone{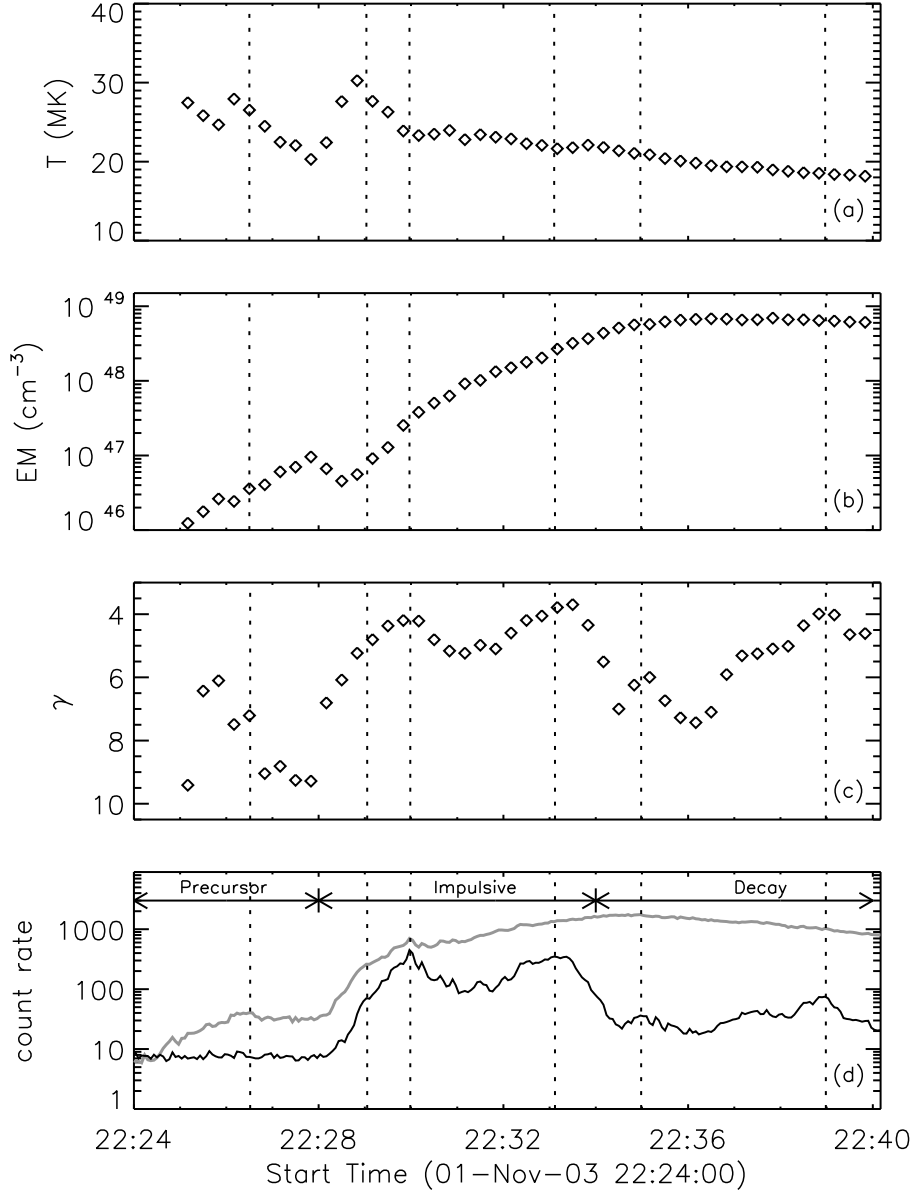}
\caption{Temporal evolution of various spectroscopic quantities derived from RHESSI X-ray spectral fits of consecutive 20 s integration time. From top to bottom: plasma temperature, emission measure, photon spectral index, and RHESSI count rates in the 6--30 keV and 30--80 keV energy bands. A few representative spectra during the important stages of the flare evolution (indicated by vertical lines) are shown in Figure \ref{rhessi_spec}.}
\label{spec_param}
\end{figure}

\begin{figure}
\epsscale{.80}
\plotone{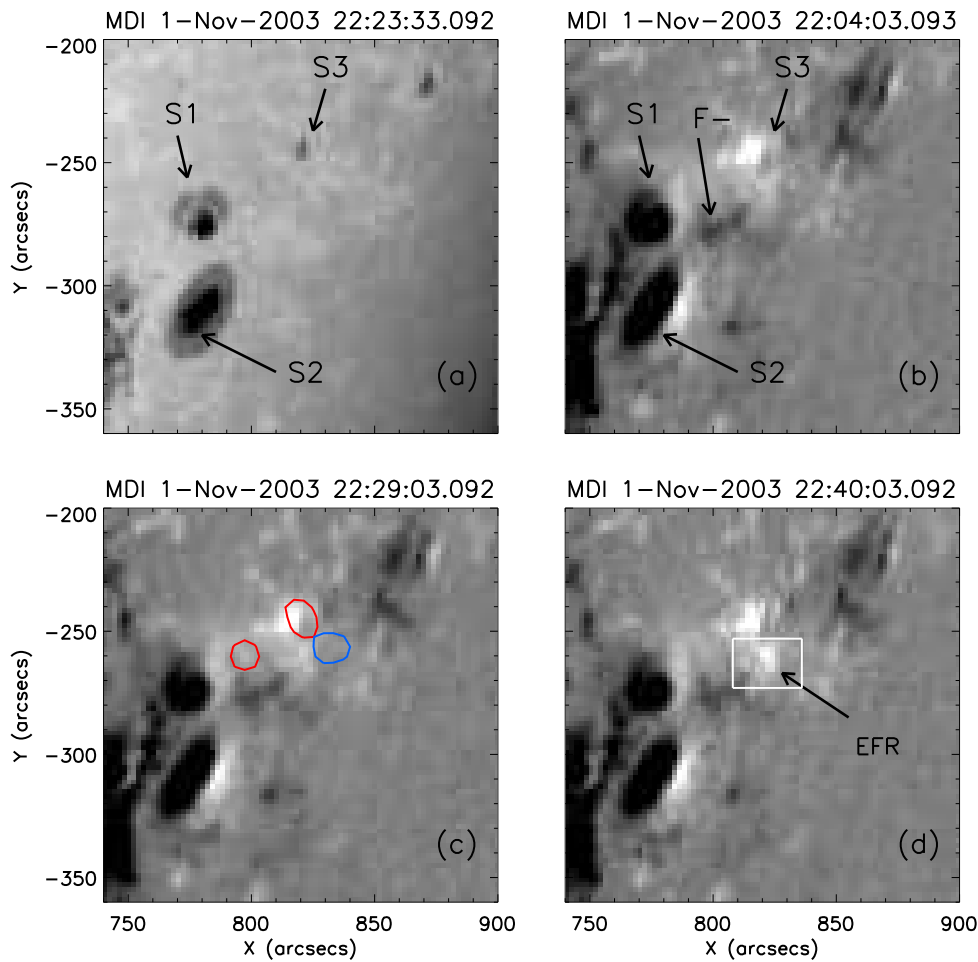}
\caption{(a) SOHO white light image close to flare onset. The three sunspots identified near the flaring region are marked by arrows and denoted as S1, S2 and S3. (b)-(d) MDI magnetograms showing magnetic field evolution during the event. The negative flux region in-between the sunspots is marked as F- in panel (b). The emerging flux region of positive polarity near the sunspot S3 is marked as EFR in panel (d).  Red and blue contours in panel (c) represent co-temporal X-ray sources at 50--100 keV and 6--12 keV energy bands respectively and denote the region where the X-ray intensity is 70\% of its peak value. Rectangular box of size 28$''$ $\times$ 20$''$ shown in panel (d) define EFR . The temporal evolution of emerging magnetic flux through this region is shown in Figure \ref{mdi_flux}.}
\label{mdi_wl_mag}   
\end{figure}

\begin{figure}
\epsscale{.80}
\plotone{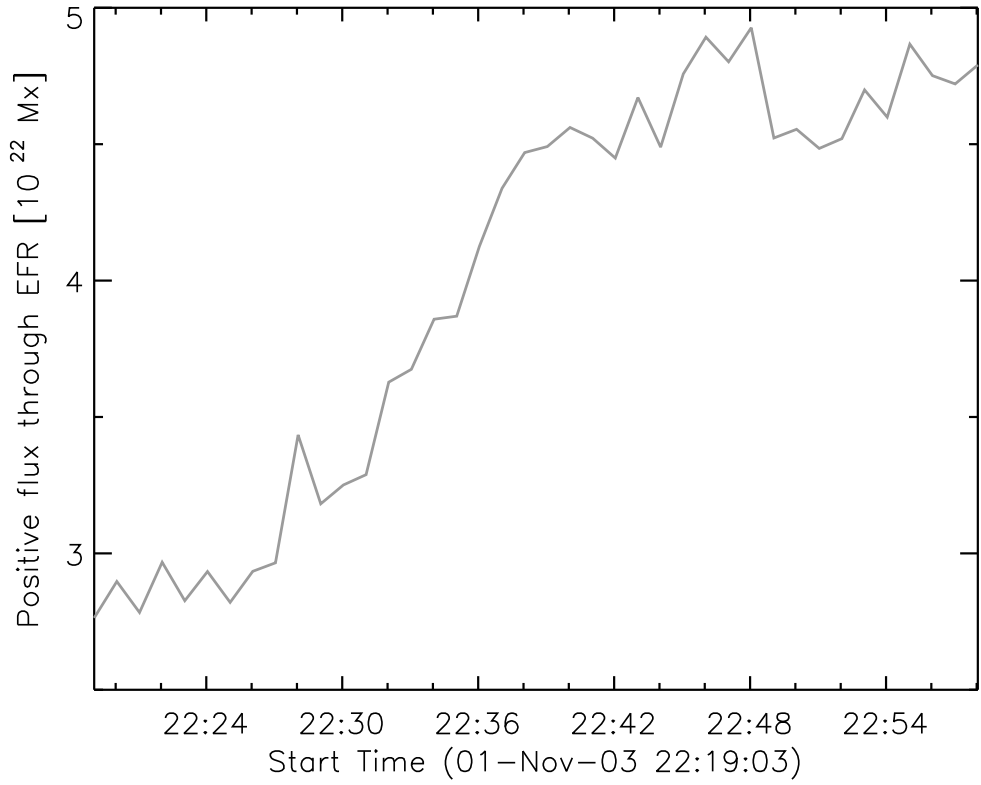}
\caption{Temporal evolution of emerging magnetic flux through EFR which is defined in Figure \ref{mdi_wl_mag}d by a rectangular box of size 28$''$ $\times$ 20$''$.}
\label{mdi_flux}   
\end{figure}

\begin{figure}
\epsscale{.80}
\plotone{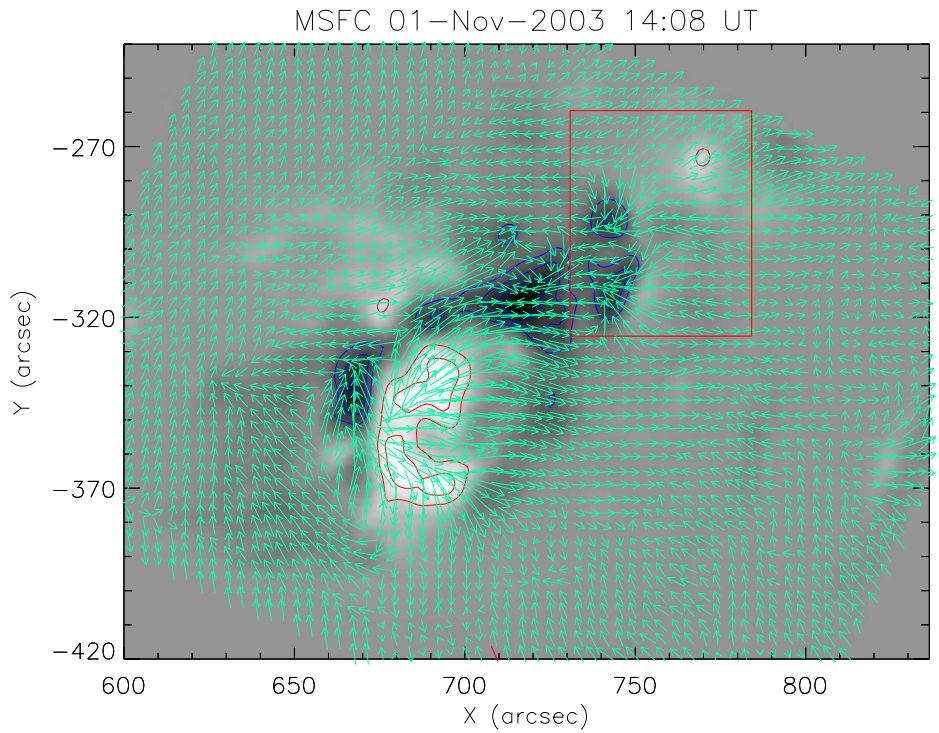}
\caption{The line of sight magnetogram of NOAA AR 10486 on November 1, 2003 at 14:08 UT overlaid with the transverse vectors (indicated by green arrows). The red/blue contours represent the positive/negative polarity. The contour levels are $\pm$ 1000, $\pm$1500, $\pm$2000 G. The red box covers the region of interest.}
\label{msfc_vector}
\end{figure}

\end{document}